\pdfoutput=1

\documentclass[11pt,twoside,a4paper,cmspaper,final,collab]{cms-tdr}
\def\svnVersion{253136}\def\cmsCernNo{2013-037}\def\cmsCernDate{\today}\def\cmsMessage{Published in the Journal of High Energy Physics as \href{http://dx.doi.org/10.1007/JHEP05(2012)055}{\doi{10.1007/JHEP05(2012)055.}}}

\begin{document}\cmsNoteHeader{EXO-11-017}

\hyphenation{had-ron-i-za-tion}
\hyphenation{cal-or-i-me-ter}
\hyphenation{de-vices}

\RCS$Revision: 183618 $
\RCS$HeadURL: svn+ssh://alverson@svn.cern.ch/reps/tdr2/papers/EXO-11-017/trunk/EXO-11-017.tex $
\RCS$Id: EXO-11-017.tex 183618 2013-04-29 19:13:56Z alverson $

\providecommand{\rd}{\ensuremath{\mathrm{d}}}
\newcommand{\CHI}{\ensuremath{\chi_\text{dijet}}\xspace}
\newcommand{\YSTAR}{\ensuremath{y^{\star}}\xspace}
\newcommand{\YBOOST}{\ensuremath{y_\text{boost}}\xspace}
\newcommand{\intlumi} {\ensuremath{2.2\fbinv}\xspace}
\newcommand{\mjj} {\ensuremath{M_{jj}}\xspace}
\newcommand{\pz}{\ensuremath{p_{\mathrm{z}}}\xspace}
\newcommand{\ptave}{\ensuremath{\langle p_{\mathrm{T}}  \rangle}\xspace}
\newcommand{\ptavens}{\ensuremath{\langle p_{\mathrm{T}}  \rangle}}
\newlength\cmsFigWidth
\ifthenelse{\boolean{cms@external}}{\setlength\cmsFigWidth{1.0\columnwidth}}{\setlength\cmsFigWidth{0.75\textwidth}}
\ifthenelse{\boolean{cms@external}}{\providecommand{\cmsLeft}{top}}{\providecommand{\cmsLeft}{left}}
\ifthenelse{\boolean{cms@external}}{\providecommand{\cmsRight}{bottom}}{\providecommand{\cmsRight}{right}}

\cmsNoteHeader{EXO-11-017} 
\title{Search for quark compositeness in dijet angular distributions from pp collisions at $\sqrt{s} = 7\TeV$}

\date{\today}

\abstract{ A search for quark compositeness using dijet angular
distributions from pp collisions at $\sqrt{s} = 7$\TeV is
presented. The search has been carried out using a data sample
corresponding to an integrated luminosity of \intlumi, recorded by
the CMS experiment at the LHC. Normalized dijet angular
distributions have been measured for dijet invariant masses from
0.4\TeV to above 3\TeV and compared with a variety of contact
interaction models, including those which take into account the
effects of next-to-leading-order QCD corrections.  The data are
found to be in agreement with the predictions of perturbative
QCD, and lower limits are obtained on the contact interaction scale, ranging
from 7.5 up to 14.5\TeV at 95\% confidence level.}

\hypersetup{%
pdfauthor={CMS Collaboration},%
pdftitle={Search for quark compositeness in dijet angular distributions from pp collisions at sqrt(s) = 7 TeV},%
pdfsubject={CMS},%
pdfkeywords={CMS, physics, dijet angular distributions, quark compositeness}}

\maketitle 

In theories of physics beyond the standard model, it has been
proposed that quarks are composite particles and are bound
states of more fundamental entities~\cite{CI-1,CI-2}. Models of
quark compositeness may explain the number of quark generations,
quark charges, and quark masses, which are not predicted in the
standard model. A common signature of quark compositeness models
is the appearance of new interactions between quark constituents
at a characteristic scale $\Lambda$ that is much larger than the
quark masses.  At energies well below $\Lambda$, these
interactions can be approximated by a contact interaction (CI)
characterized by a four-fermion coupling. In this Letter,
flavor-diagonal color-singlet couplings between quarks are
studied. These can be described by the effective
Lagrangian~\cite{CI-1,bib_PDG}

\begin{equation*}
L_{\cPq\cPq}=\frac{2\pi}{\Lambda^2} \left[
\eta_{LL}({\overline{\cPq}}_{L}\gamma^{\mu}\cPq_{L})(\overline{\cPq}_{L}\gamma_{\mu}\cPq_{L})
+\eta_{RR}(\overline{\cPq}_{R}\gamma^{\mu}\cPq_{R})(\overline{\cPq}_{R}\gamma_{\mu}\cPq_{R})
+2\eta_{RL}(\overline{\cPq}_{R}\gamma^{\mu}\cPq_{R})(\overline{\cPq}_{L}\gamma_{\mu}\cPq_{L})
\right],
\end{equation*}

where the subscripts $L$ and $R$ refer to the chiral projections of the quark fields and
$\eta_{LL}$, $\eta_{RR}$, and $\eta_{RL}$ can be 0, ${+}1$, or ${-}1$.
The various combinations of $\eta_{LL}$, $\eta_{RR}$, and
$\eta_{RL}$ correspond to different CI models. The following CI
scenarios are investigated:

\begin{align*}
\Lambda &= \Lambda^{\pm}_{LL}~{\rm for}~(\eta_{LL}, \eta_{RR}, \eta_{RL}) = (\pm 1, 0, 0) ,\\
\Lambda &= \Lambda^{\pm}_{RR}~{\rm for}~(\eta_{LL}, \eta_{RR}, \eta_{RL}) = (0, \pm 1, 0) ,\\
\Lambda &= \Lambda^{\pm}_{VV}~{\rm for}~(\eta_{LL}, \eta_{RR}, \eta_{RL}) = (\pm 1, \pm 1, \pm 1) ,\\
\Lambda &= \Lambda^{\pm}_{AA}~{\rm for}~(\eta_{LL}, \eta_{RR}, \eta_{RL}) = (\pm 1, \pm 1, \mp 1) ,\\
\Lambda &= \Lambda^{\pm}_{(V-A)}~{\rm for}~(\eta_{LL}, \eta_{RR}, \eta_{RL}) = (0, 0, \pm 1) .
\end{align*}

In pp collisions these models result in the same limits for $\Lambda^{\pm}_{LL}$ and
$\Lambda^{\pm}_{RR}$, and at tree level for $\Lambda^{\pm}_{VV}$ and
$\Lambda^{\pm}_{AA}$ as well as for $\Lambda^{+}_{(V-A)}$ and $\Lambda^{-}_{(V-A)}$.

High energy proton-proton collisions with large momentum transfers
predominantly produce events containing two jets with high
transverse momenta (dijets). Such events probe the scattering
partons at the shortest distance scales and provide a fundamental
test of quantum chromodynamics (QCD). The angular distribution of
these two jets with respect to the beam direction is directly
sensitive to the underlying dynamics of the parton-parton
scattering and does not strongly depend on the parton distribution
functions (PDFs). Distributions of the polar scattering angle
$\theta^{*}$ in the parton-parton center-of-mass frame from QCD
processes are peaked in the forward and backward directions,
whereas contact interactions give rise to more isotropic
distributions in $\theta^{*}$.

Previous searches for quark compositeness at hadron colliders
have been reported at
the S{\Pp\Pap}S by the UA1~\cite{bib_UA1} collaboration, at
the Tevatron by the D0~\cite{bib_d0-prl-chi,bib_d0-prd-chi} and
CDF~\cite{bib_cdf-prl-chi} collaborations, and at the Large Hadron
Collider (LHC) by the ATLAS~\cite{bib_atlas-dijet,bib_atlas-chi}
and CMS~\cite{bib_cms-chi2010,bib_cms-dijets} collaborations. The
limits on quark compositeness at the LHC~\cite{bib_cms-chi2010,
bib_cms-dijets, bib_atlas-dijet, bib_atlas-chi} have been reported
only for a color- and isospin-singlet CI model,
$\Lambda^{\pm}_{LL/RR}$~, where $\Lambda^{+}_{LL/RR}
(\Lambda^{-}_{LL/RR})$ corresponds to destructive (constructive)
interference between the CI and QCD terms. In this Letter, our
previous searches are extended to higher CI scales using a data
sample corresponding to an integrated luminosity of \intlumi at
$\sqrt s$ = 7\TeV, exploring for the first time at the LHC
a wide range of CI models. Also, this is the first use of
a recent CI prediction that includes next-to-leading-order
(NLO) QCD corrections~\cite{gao_nlo}.

In this analysis, the normalized dijet angular distributions,
defined as
$(1/\sigma_\text{dijet})(\rd\sigma_\text{dijet}/\rd\chi_\text{dijet})$
where $\chi_\text{dijet} = \mathrm{e}^{|y_1 -y_2|}$, are studied for
several ranges of the dijet invariant mass \mjj. Here, $y_1$ and
$y_2$ are the rapidities of the two highest transverse momentum
(\PT) jets, and they are related to the jet energy $E$ and the
projection of the jet momentum on the beam axis, \pz, by $y =
{\frac{1}{2}} \ln [(E + \pz) / (E - \pz)]$. In the limit of
massless scattering partons, $\chi_\text{dijet}$ is related to
$\theta^{*}$ by $\chi_\text{dijet} = (1 + |\cos{\theta^*}|) / ( 1
- |\cos{\theta^*}|)$. The use of the variable $\chi_\text{dijet}$
is motivated by the fact that
$\rd\sigma_\text{dijet}/\rd\chi_\text{dijet}$ is approximately uniform
for QCD dijet processes, while CI models predict angular distributions
that are strongly peaked at low values of $\chi_\text{dijet}$.

The data for this analysis are collected with the Compact Muon
Solenoid (CMS) detector at the LHC. The central feature of the CMS
detector is a superconducting solenoid, 12.5\unit{m} long and with an
internal diameter of 6\unit{m}, providing an axial field of 3.8\unit{T}. The
field volume of the solenoid is instrumented with various particle
detection systems. Charged particle trajectories are measured by a
silicon pixel and strip tracker, covering $0<\varphi<2\pi$ in
azimuth and $|\eta| <2.5$, where pseudorapidity $\eta =
-\text{ln} [\text{tan}(\theta/2)]$ and $\theta$ is the polar angle
relative to the counterclockwise proton beam direction with
respect to the center of the detector. A lead-tungstate crystal
electromagnetic calorimeter and a brass/scintillator hadronic
calorimeter surround the tracking volume. A preshower detector
made of silicon sensor planes and lead absorbers is installed in
front of the electromagnetic calorimeter at $1.653 <\vert \eta
\vert < 2.6$. Outside the solenoid, muons are measured in
gas-ionization detectors embedded in the steel return yoke. A more
detailed description of the CMS detector can be found
elsewhere~\cite{bib_CMS}.

The CMS detector records events with a two-tiered trigger system
consisting of a hardware-based Level-1 (L1) and a software-based
High Level Trigger (HLT). In this study, single-jet triggers that
reconstruct jets from calorimeter energy deposits at L1 and HLT
are used to select events based on different jet-\PT thresholds.
Seven combinations of (L1, HLT) \PT thresholds (in GeV) are used
to select events: (36, 60), (68, 80), (92, 110), (92, 150), (92,
190), (92, 240), and (92, 300).  All except the highest-threshold
jet trigger were prescaled during the 2011 run. The
efficiency of each single-jet trigger is measured as a function of
\mjj using events selected by a lower-threshold trigger.

Jets are reconstructed offline using the anti-$k_T$ clustering
algorithm with a distance parameter $R = 0.5$~\cite{bib_antikt}.
The four-vectors of particles reconstructed by the CMS
particle-flow algorithm are used as input to the jet-clustering
algorithm. The particle-flow algorithm~\cite{bib_pft09,bib_pft10}
combines information from all CMS subdetectors to provide a
complete list of long-lived particles in the event. Reconstructed
and identified particles include muons, electrons (with associated
bremsstrahlung photons), photons (including conversions in the
tracker volume), and charged and neutral hadrons. The jet energy
scale is calibrated using measurements of energy balance in dijet
and photon+jet events~\cite{bib_jecjinst}. Extra energy clustered
into jets from additional proton-proton interactions within the
same bunch crossing (pileup) is taken into account on an event-by-event
basis by a correction
to the jet four-vectors. The average number of pileup interactions
for the data sample used in this analysis has been estimated to be 5.

Events with at least two reconstructed jets are selected from an 
inclusive non flavored-tagged jet sample, and the two highest-\PT 
jets are used to measure the dijet angular
distributions for different ranges in \mjj. Events with spurious
jets from noise and non-collision backgrounds are rejected by
applying loose quality criteria to the jet
properties~\cite{bib_newjec} and requiring a reconstructed
primary vertex within $\pm 24\unit{cm}$ of the detector center along the
beam line and within 2 cm of the detector center in the plane
transverse to the beam~\cite{bib_vertexpas}. The rapidities
$|y_1|$ and $|y_2|$ of the two highest-\PT jets are restricted
to be less than 2.5 by selecting only events with
$\chi_\text{dijet}<16$ and $|y_\text{boost}|<1.11$, where $y_\text{
boost} = {\frac{1}{2}}(y_1 + y_2)$.
The lower limits of the $M_{jj}$ ranges for the dijet angular
distributions were chosen such that the trigger efficiencies
exceed $99\%$, and are given by the values $0.4, 0.6, 0.8, 1.0,
1.2, 1.5, 1.9, 2.4,$ and $3.0$\TeV. The data for the first five
$M_{jj}$ ranges are recorded using prescaled triggers and
correspond to integrated luminosities of 0.77, 5.9, 32, 108, and
371~pb$^{-1}$, while the data for the mass ranges with $M_{jj} >
1.5$\TeV correspond to the full integrated luminosity of \intlumi.
The uncertainty on the integrated luminosities has been estimated
to be 4.5\%~\cite{bib_lumi2011,bib_lumi}. The highest value of $M_{jj}$
observed in this data sample is 3.9\TeV.

The dijet angular distributions are corrected for migration
effects due to the finite jet energy and position resolutions. The
four-momenta, rapidities, and azimuthal angles of generated jets
from Monte Carlo (MC) event simulations are varied within their
measured resolutions~\cite{bib_jecjinst}, and correction factors
for each $M_{jj}$ region are obtained from the ratio of the
generated to the smeared $\chi_\text{dijet}$ distributions.
Unfolding correction factors are evaluated from two independent MC
samples, \PYTHIA 6.422~\cite{bib_pythia} with tune \textsc{
D6T}~\cite{bib_pythiatunes} and {\HERWIG{}++}
2.4.2~\cite{bib_herwig} with tune \textsc{2.3}, and the average of
these corrections is applied to the data. The size of the
correction factors varies from less than 1.3\% in the lowest
$M_{jj}$ range to less than 10\% in the highest $M_{jj}$ range.
The associated systematic uncertainties are taken as the maximum
differences between the unfolding  corrections obtained from four
independent MC samples, \HERWIG{}++ tune 2.3, \PYTHIA{}6 tunes D6T
and Z2 (the Z2 tune is identical to the Z1 tune~\cite{bib_pythiatunes}
except that Z2 uses the CTEQ6L PDF~\cite{cteq6l}), and \PYTHIA{}8~\cite{bib_pythia8} tune
4C~\cite{bib_tune4c}, and the nominal correction factors. These
uncertainties range from less than 0.2\% at low $M_{jj}$ to less
than 4.9\% at high $M_{jj}$. A systematic uncertainty from using a
parameterized model to simulate the finite jet \PT and position
resolutions to determine the unfolding correction factors is
estimated by comparing the smeared $\chi_\text{dijet}$
distributions to the ones from a detailed simulation of the CMS
detector using \GEANTfour~\cite{GEANT}.  This uncertainty is
found to be less than 1.3\% (2.0\%) in the lowest (highest) $M_{jj}$ range
and is added in quadrature to the unfolding uncertainties.

The dijet angular distributions are normalized to the integrated
dijet cross sections in each $M_{jj}$ range and are relatively
insensitive to many systematic effects. For example, they show
little dependence on the overall jet-energy scale and are
independent of the luminosity uncertainty. However, they are
sensitive to the rapidity dependence of the jet energy calibration
and to the jet \PT resolution. For the phase space in \PT and
$\eta$ of the jets in this analysis, the jet energy scale
uncertainties vary between 2\% and 3\% and have a dependence on
pseudorapidity of less than 1\% per unit of
$\eta$~\cite{bib_jecjinst}. The uncertainty on the jet \PT
resolution is less than 10\%~\cite{bib_jecjinst}. The resulting
uncertainty on the $\chi_\text{dijet}$ distributions due to the
jet energy calibration uncertainties is found to be less than
1.0\% at low $M_{jj}$ and less than 0.3\% at high $M_{jj}$ over
all $\chi_\text{dijet}$ bins, while
the maximum uncertainty due to the jet \PT resolution uncertainty varies
from 0.2\% at low $M_{jj}$ to 0.6\% at high $M_{jj}$. In addition,
uncertainties on the tails of the jet \PT
resolutions~\cite{bib_jecjinst} result in systematic uncertainties
on the $\chi_\text{dijet}$ distributions ranging from less than
0.5\% at low $M_{jj}$ to less than 4.6\% at high $M_{jj}$. The
effect of pileup was investigated by dividing the data into low
and high pileup samples based upon the vertex multiplicity, and
comparing the $\chi_\text{dijet}$ distributions from each sample.
No significant effect was observed. The
total systematic uncertainty on the $\chi_\text{dijet}$
distributions, calculated as the quadratic sum of the
contributions due to the uncertainties in the jet energy
calibration, the jet \PT resolution, and the unfolding
correction, is less than 1.7\% for the lowest $M_{jj}$ range and
less than 7\% for the highest $M_{jj}$ range. A summary of the
leading systematic uncertainties is provided in
Table~\ref{tab:sys}.

\begin{table*}[htbp]
\begin{center}
\caption{Summary of the leading experimental and theoretical
uncertainties on the \CHI distributions. The maximum uncertainties
over all \CHI bins for the lowest and highest $M_{jj}$ ranges are
given. The dominant experimental contribution is the statistical
uncertainty while the dominant contribution to the NLO
perturbative QCD uncertainty is the factorization and
renormalization scale uncertainty.} \label{tab:sys}
\begin{tabular}{l|c|c}
\hline
\multirow{2}{*}{\centering Source of Uncertainty} & $0.4<M_{jj}<0.6$\TeV & $M_{jj}>3$\TeV \\
                       &    (\%)     &  (\%) \\
\hline
Jet energy scale & 1.0 & 0.3 \\
Jet energy resolution & 0.2 & 0.6 \\
Jet energy resolution tails & 0.5 & 4.6 \\
Unfolding, MC modeling & 0.2 & 4.9 \\
Unfolding, detector simulation & 1.3 & 2.0 \\
\hline
Total experimental systematic uncertainty & 1.7 & 7.0 \\
\hline
Statistical uncertainty & 2.5 & 31.6 \\
\hline
$\mu_r$ and $\mu_f$ scales & 5.6 & 14.9 \\
PDF (CTEQ6.6) & 0.5 & 0.7 \\
Non-perturbative corrections & 1.7 & 1.1 \\
\hline
Total theoretical systematic uncertainty & 5.9 & 15.0 \\
\hline
\end{tabular}
\end{center}
\end{table*}

Predictions at NLO in perturbative QCD are made for the dijet angular
distributions with \textsc{nlojet++} 2.0.1~\cite{bib_nlojet} in the
\textsc{fastnlo} framework version 1.4~\cite{fastnlo}. The factorization
($\mu_{f}$) and renormalization ($\mu_{r}$) scales are set to \ptave,
the mean \PT of the two jets, and the PDFs are taken from the
CTEQ6.6 set~\cite{cteq}. Correction factors are applied to the
predictions to account for non-perturbative effects due to
hadronization and multiple parton interactions. These correction
factors are used to correct the parton QCD calculations to the
particle level, and they are determined by the average of the
corrections estimated using \PYTHIA{}6 tune {D6T} and {\HERWIG{}++} tune 2.3.
This uncertainty is found to encompass
alternative choices of MC tunes, \PYTHIA{}6 tune Z2 or \PYTHIA{}8 tune 4C, and is estimated to be less than 1.7\% (1.1\%)
at low (high) $M_{jj}$.

The dominant source of uncertainty on the QCD predictions is due
to the choices of the $\mu_{f}$ and $\mu_{r}$ scales. The
uncertainty is evaluated following the proposal in
Ref.~\cite{Banfi:2010xy} by varying the default choice of scales
in the following 6 combinations:
$(\mu_{f}/\ptave$, $\mu_{r}/\ptave)$ =
$(1/2, 1/2)$, $(1/2, 1)$, $(1,1/2)$, $(2, 2)$, $(2, 1)$ and $(1, 2)$.
These scale variations modify the predictions of the normalized
$\chi_\text{dijet}$ distributions by less than 5.6\% (15\%) at low
(high) $M_{jj}$. The uncertainty due to the choice of PDFs is
determined from the 22 uncertainty eigenvectors of CTEQ6.6 using the
procedure described in Ref.~\cite{cteq}, and is found to be less
than 0.5\% at low $M_{jj}$ and less than 0.7\% at high $M_{jj}$.
The leading systematic uncertainties on the
theoretical predictions are listed in Table~\ref{tab:sys}.

\begin{figure}[htbp]
\centering
\includegraphics[width=\cmsFigWidth]{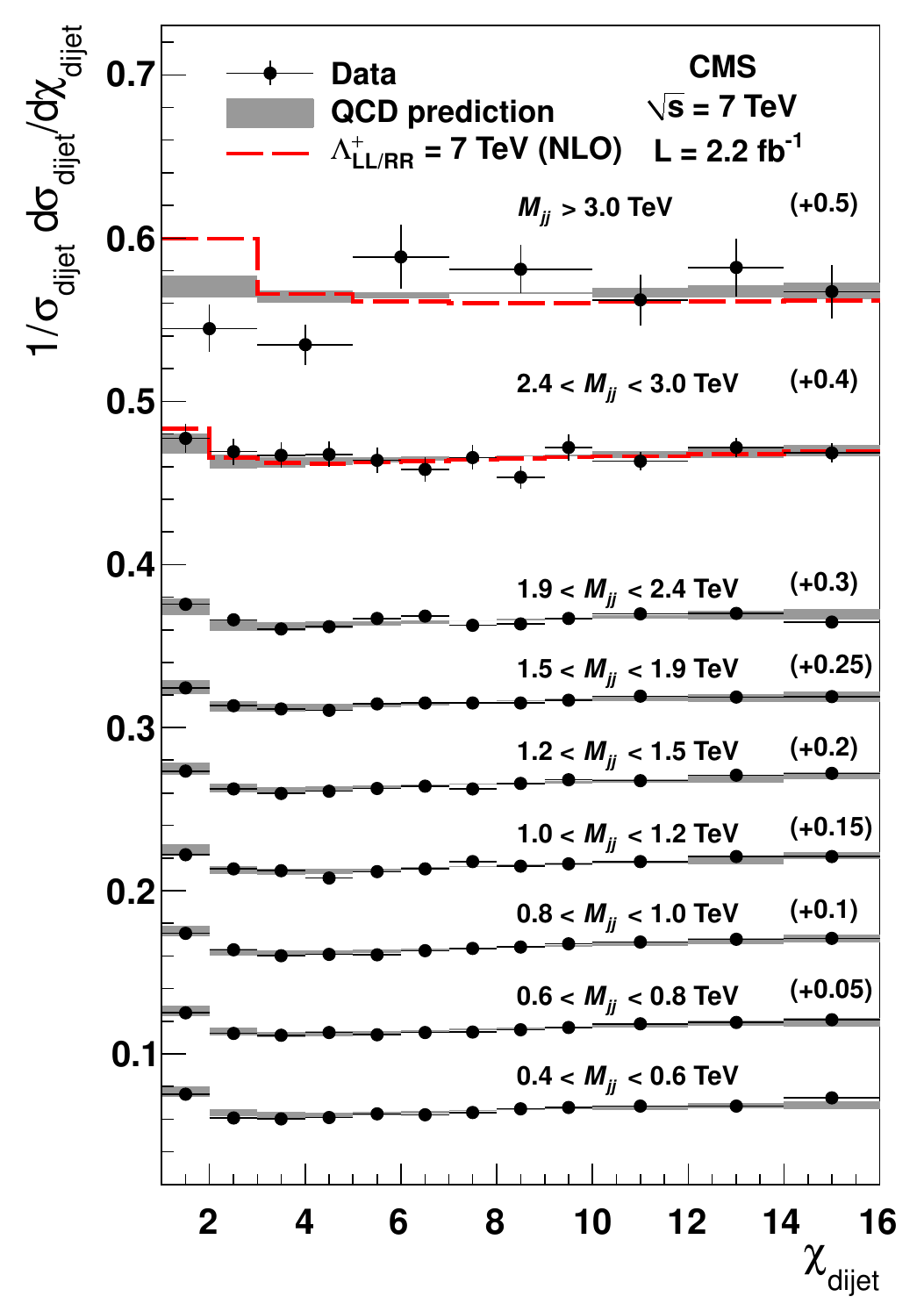}
\caption{\label{fig:chi}Normalized dijet angular distributions for
  $|y_\text{boost}|<1.11$ in several $M_{jj}$ ranges. For clarity, the
  distributions are shifted vertically by the
  additive amounts shown in parentheses in the figure.
  The vertical error bars represent the statistical and systematic uncertainties added in
  quadrature. The horizontal bars correspond to the $\chi_\text{dijet}$ bin width. The results are
  compared with the predictions of NLO QCD with CTEQ6.6 PDF
  (shaded band) and with predictions for
  QCD+CI from~\cite{gao_nlo} at the CI scale
  $\Lambda^{+}_{LL/RR}=7$\TeV (dashed histogram). Non-perturbative
  corrections due to hadronization and multiple parton interactions
  are applied to the predictions. The shaded band
  indicates the total uncertainty on the NLO QCD predictions due to
  $\mu_r$ and $\mu_f$ scale variations, PDFs, as well as the
  uncertainties from the non-perturbative corrections, which have all
  been added in quadrature.}
\end{figure}

The measured differential dijet angular distributions, corrected
for instrumental effects and normalized to their respective
integrals, are compared to QCD predictions in Fig.~\ref{fig:chi}
for different $M_{jj}$ ranges. Overall the theoretical predictions
provide a good description of the data for all $M_{jj}$ ranges.

The measured dijet angular distributions are used to set limits on
a variety of CI models. Only color-singlet models, which predict
the largest deviations of the dijet angular distributions from the
standard model, are considered. In fact, for the general case of a
CI model containing both color-singlet and color-octet
contributions, there are certain regions in the theory parameter
space where the CI predictions for the dijet angular distributions
become indistinguishable from the QCD predictions, because of
interference between these contributions~\cite{gao_nlo}.

\begin{figure}[htbp]
\centering
\includegraphics[width=\cmsFigWidth]{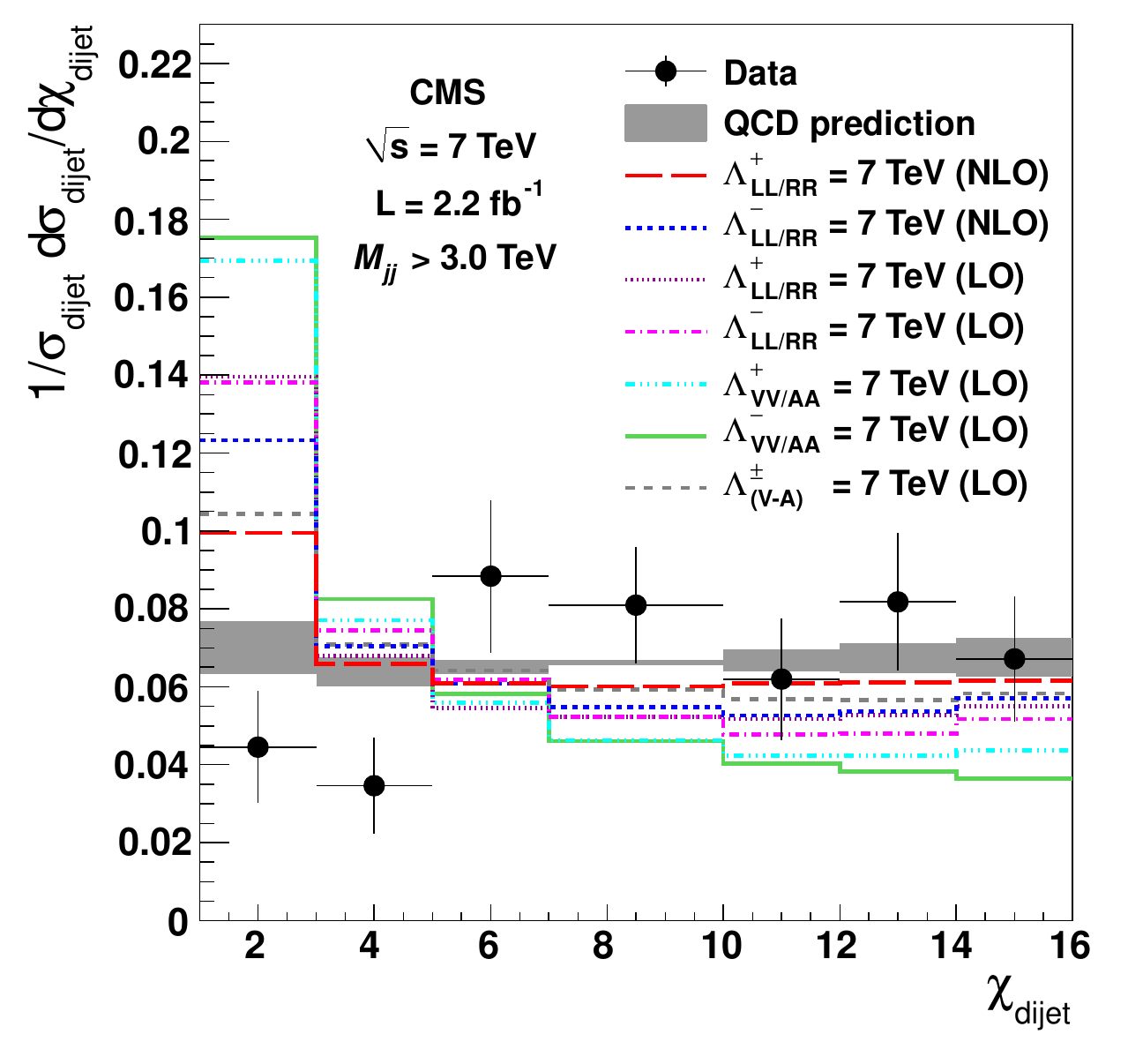}
\caption{\label{fig:theories}The normalized dijet angular
distribution in the highest dijet mass range $M_{jj}>3$\TeV
compared to various contact interaction models (dashed and dotted
histograms). Non-perturbative corrections due to hadronization and
multiple parton interactions are applied to the predictions. The
vertical error bars represent the statistical and systematic
uncertainties added in quadrature. The horizontal bars
correspond to the $\chi_\text{dijet}$ bin width.  The shaded band
represents the NLO QCD prediction and includes the systematic
uncertainties due to $\mu_r$ and $\mu_f$ scale variations and PDF
uncertainties, as well as the uncertainties from the
non-perturbative corrections added in quadrature.}
\end{figure}

In this analysis we present limits for a CI model that includes
the exact NLO QCD corrections to dijet production induced by
contact interactions~\cite{gao_nlo}, as well as limits extracted
from various CI models implemented in \PYTHIA{}8~\cite{bib_pythia8}. In the latter case, the contributions
of CI and QCD are calculated to leading order (LO).
To take into account the NLO QCD corrections which are missing in
the \PYTHIA{}8 model, the cross-section difference
$\sigma^{\text{QCD}}_{\text{NLO}} -
\sigma^{\text{QCD}}_{\text{LO}}$ is added to the LO QCD+CI
prediction in each $M_{jj}$ and $\chi_\text{dijet}$ bin. With this
procedure, we obtain a QCD+CI prediction where the QCD terms are
corrected to NLO while the CI terms are calculated at LO.
Non-perturbative corrections due to hadronization and multiple
parton interactions are applied to the predictions. In
Fig.~\ref{fig:chi}, the predictions are shown for QCD+CI
from Ref.~\cite{gao_nlo} at the CI scale $\Lambda^{+}_{LL/RR}=7$\TeV
for the two highest \mjj ranges. The highest \mjj range is the
most sensitive to the CI signal, though omitting the
second-highest \mjj range would decrease the expected limit on the
CI scale by about 1\% for $\Lambda^{+}_{LL/RR}$ and 13\% for
$\Lambda^{-}_{LL/RR}$. Varying the boundary between the two
highest \mjj ranges by $\pm 0.2$\TeV changes the expected limit by
less then 0.5\%. The predictions for the various QCD+CI models at
the scale of $\Lambda = 7$\TeV are shown in
Fig.~\ref{fig:theories} for the highest \mjj range. At low
$\chi_\text{dijet}$, the CI predictions with exact NLO QCD
corrections show smaller enhancement relative to QCD than the
corresponding LO CI predictions, as described in detail
in Ref.~\cite{gao_nlo}.

\begin{figure}[htbp]
\centering
\includegraphics[width=\cmsFigWidth]{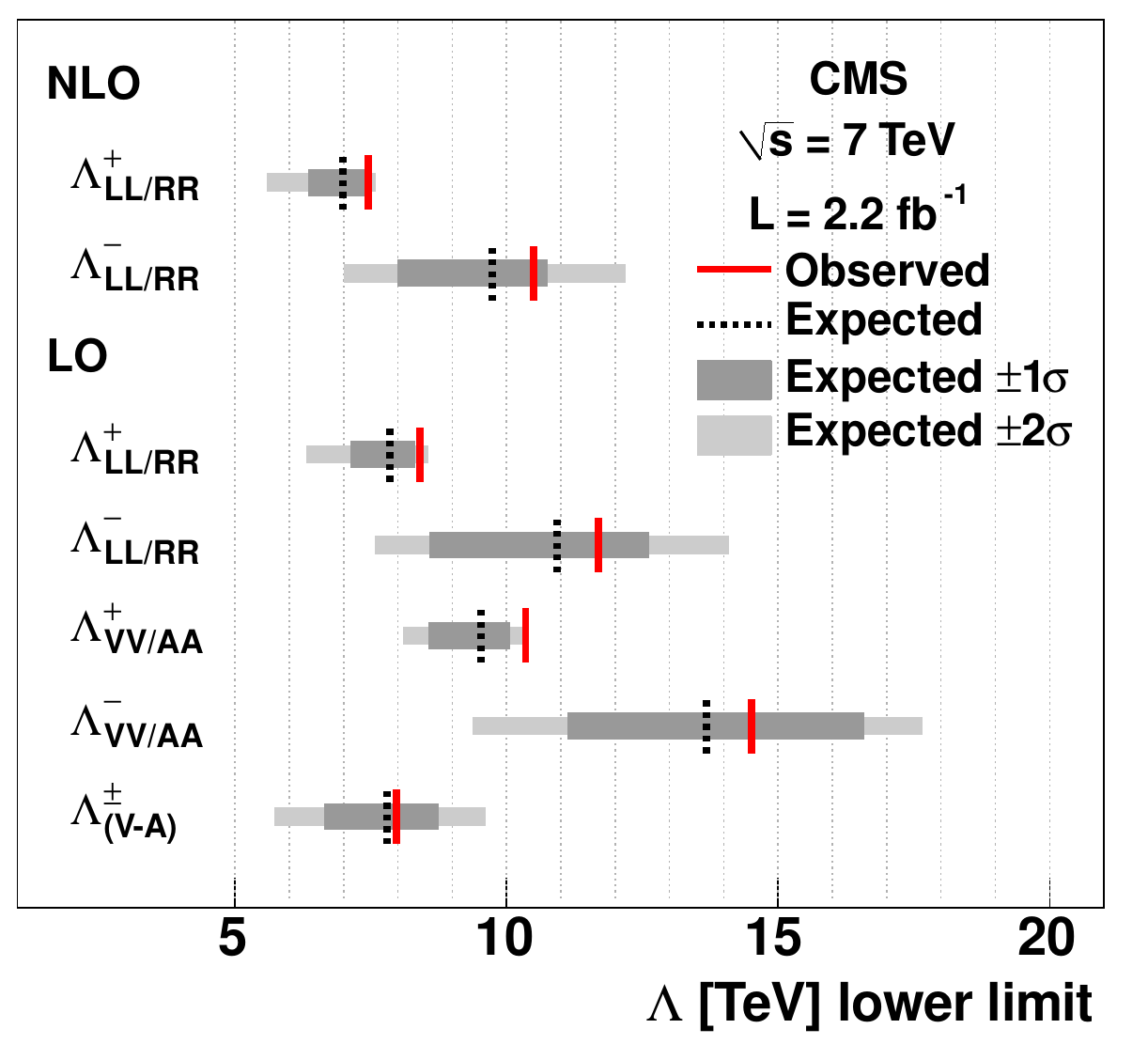}
\caption{\label{fig:limits}Observed (solid lines) and expected
(dashed lines) 95\% C.L. limits for the contact
interaction scale $\Lambda$ for the different CI models. The dark (light) gray
bands indicate the $\pm 1 \sigma$ ($\pm 2 \sigma$) uncertainties on the expected limits.}
\end{figure}

The statistical method used to set limits on $\Lambda$ follows a
modified frequentist approach~\cite{junk,read,read_physg,bib_PDG}.
The log-likelihood-ratio $q = -2 \ln
(\frac{L_\text{QCD+CI}}{L_\text{QCD}})$ is used to discriminate
between the QCD-only hypothesis and the QCD+CI hypothesis. The
$L_\text{QCD+CI}$ and $L_\text{QCD}$ are written as a product of
Poissonian likelihood functions for each bin in
$\chi_\text{dijet}$ and for the two highest ranges of \mjj, where
the predictions for each \mjj range are normalized to the number
of observed events in that range. The p-values, $P_\text{QCD+CI}(q
\geq q_\text{obs})$ and $P_\text{QCD}(q \leq q_\text{obs})$, are
obtained from ensembles of pseudo-experiments for the two
hypotheses. Systematic uncertainties are represented by nuisance
parameters which affect the $\chi_\text{dijet}$ distribution. The
nuisance parameters are varied within their Gaussian uncertainties
when generating the distributions of $q$. The QCD+CI model is
considered to be excluded at the 95\% confidence level (C.L.)
based on the quantity
${\rm CL}_{\rm s} = P_\text{QCD+CI}(q\geq q_\text{obs}) /
(1-P_\text{QCD}(q \leq q_\text{obs})) < 0.05$.
The observed and expected limits at 95\% C.L. for the CI models
considered are listed in Table~\ref{tab:limits} and displayed in
Fig.~\ref{fig:limits}. All the observed limits agree within
uncertainties with the expected limits, which are evaluated at the
median of the test statistics distribution of the QCD-only model.
The observed limits are slightly higher than the expected limits
because, for the range $M_{jj} > 3.0$\TeV, the measured dijet
angular distribution at low $\chi_\text{dijet}$ is lower than,
although statistically compatible with, the QCD prediction. The
limits for the CI scale $\Lambda^{+}_{LL/RR}$ are also extracted
using an alternative procedure in which the data are not corrected
for detector effects and instead the MC predictions are convoluted
with the detector resolutions. The limits obtained are found to agree
with the quoted ones within 1.5\%.

\begin{table}[htbp]
\begin{center}
\renewcommand{\arraystretch}{1.3}
\caption{Observed and expected lower limits at 95\% confidence
level for the contact interaction scale $\Lambda$ for several
quark CI models.} \label{tab:limits}
\begin{tabular}{l|c|c}
\hline
CI model & Observed limit (\TeVns) & Expected limit (\TeVns)\\
\hline
NLO $  \Lambda_{LL/RR}^{+}$ & 7.5 & 7.0$^{+0.4}_{-0.6}$\\
NLO $  \Lambda_{LL/RR}^{-}$ & 10.5 & 9.7$^{+1.0}_{-1.7}$\\
LO $  \Lambda_{LL/RR}^{+}$ & 8.4 & 7.9$^{+0.5}_{-0.7}$\\
LO $  \Lambda_{LL/RR}^{-}$ & 11.7 & 10.9$^{+1.7}_{-2.4}$\\
LO $  \Lambda_{VV/AA}^{+}$ & 10.4 & 9.5$^{+0.5}_{-1.0}$\\
LO $  \Lambda_{VV/AA}^{-}$ & 14.5 & 13.7$^{+2.9}_{-2.6}$\\
LO $  \Lambda_{(V-A)}^{\pm}$ & 8.0 & 7.8$^{+1.0}_{-1.1}$\\
\hline
\end{tabular}
\end{center}
\end{table}

In summary, normalized dijet angular distributions have been measured
with the CMS detector over a wide range of dijet invariant masses. The
distributions are found to be in agreement with predictions
of QCD and are used to set lower limits on the contact interaction scale for a
variety of quark compositeness models, including models with NLO
QCD corrections. The 95\% confidence level lower limits for the
contact interaction scale $\Lambda$ are in the range $7.5 -
14.5$\TeV. These results represent the most comprehensive set of limits on
the contact interaction scale to date.

We would like to thank J.~Gao, C.-S.~Li, J.~Wang, C.-P.~Yuan, and
H.-X.~Zhu for useful discussions and for providing the program to
calculate the contact interaction predictions with NLO QCD corrections.
We congratulate our colleagues in the CERN accelerator departments for
the excellent performance of the LHC machine. We thank the technical
and administrative staff at CERN and other CMS institutes, and
acknowledge support from: FMSR (Austria); FNRS and FWO (Belgium);
CNPq, CAPES, FAPERJ, and FAPESP (Brazil); MES (Bulgaria); CERN; CAS,
MoST, and NSFC (China); COLCIENCIAS (Colombia); MSES (Croatia); RPF
(Cyprus); MoER, SF0690030s09 and ERDF (Estonia); Academy of Finland,
MEC, and HIP (Finland); CEA and CNRS/IN2P3 (France); BMBF, DFG, and
HGF (Germany); GSRT (Greece); OTKA and NKTH (Hungary); DAE and DST
(India); IPM (Iran); SFI (Ireland); INFN (Italy); NRF and WCU (Korea);
LAS (Lithuania); CINVESTAV, CONACYT, SEP, and UASLP-FAI (Mexico); MSI
(New Zealand); PAEC (Pakistan); MSHE and NSC (Poland); FCT (Portugal);
JINR (Armenia, Belarus, Georgia, Ukraine, Uzbekistan); MON, RosAtom,
RAS and RFBR (Russia); MSTD (Serbia); MICINN and CPAN (Spain); Swiss
Funding Agencies (Switzerland); NSC (Taipei); TUBITAK and TAEK
(Turkey); STFC (United Kingdom); DOE and NSF (USA).

\bibliography{auto_generated}   
\cleardoublepage \appendix\section{The CMS Collaboration \label{app:collab}}\begin{sloppypar}\hyphenpenalty=5000\widowpenalty=500\clubpenalty=5000\textbf{Yerevan Physics Institute,  Yerevan,  Armenia}\\*[0pt]
S.~Chatrchyan, V.~Khachatryan, A.M.~Sirunyan, A.~Tumasyan
\vskip\cmsinstskip
\textbf{Institut f\"{u}r Hochenergiephysik der OeAW,  Wien,  Austria}\\*[0pt]
W.~Adam, T.~Bergauer, M.~Dragicevic, J.~Er\"{o}, C.~Fabjan, M.~Friedl, R.~Fr\"{u}hwirth, V.M.~Ghete, J.~Hammer\cmsAuthorMark{1}, M.~Hoch, N.~H\"{o}rmann, J.~Hrubec, M.~Jeitler, W.~Kiesenhofer, M.~Krammer, D.~Liko, I.~Mikulec, M.~Pernicka$^{\textrm{\dag}}$, B.~Rahbaran, C.~Rohringer, H.~Rohringer, R.~Sch\"{o}fbeck, J.~Strauss, A.~Taurok, F.~Teischinger, P.~Wagner, W.~Waltenberger, G.~Walzel, E.~Widl, C.-E.~Wulz
% ...
\vskip\cmsinstskip
\dag:~Deceased\\
1:~~Also at CERN, European Organization for Nuclear Research, Geneva, Switzerland\\
% ...

\end{sloppypar}
\end{document}